\documentclass[sigplan,]{acmart}

\usepackage{booktabs} 
\usepackage{multirow}
\usepackage{etoolbox,siunitx}
\usepackage[utf8]{inputenc}
\usepackage{microtype}
\usepackage[]{todonotes}
\usepackage{balance}
\usepackage{subcaption}
\usepackage{tikz}
\usepackage{cleveref}

\usepackage{etoolbox}
\newcommand{\ubold}{\fontseries{b}\selectfont} 
\robustify\ubold

\usetikzlibrary{shapes, arrows, positioning, backgrounds}
\tikzstyle{kb} = [cylinder, text=black, text centered, shape border rotate=90, aspect=0.25, draw, align=center, text
width = 6.6em]
\tikzstyle{prg} = [rectangle, rounded corners, text=black, text centered, draw, align=left, inner sep=5pt]
\tikzstyle{select} = [circle, text=black, text centered, draw, align=left, inner sep=0pt]
\tikzstyle{txt} = [rectangle, text=black, text centered, draw=white, align=left, text width = 2.2em]
\tikzstyle{txts} = [rectangle, text=black, text centered, draw=white, align=left, text width = 1em]
\tikzstyle{line} = [->, draw]
\tikzstyle{doubleline} = [<->, draw]

\newcommand*\circled[1]{\tikz[baseline= (char.base)]{
  \node[shape=circle,draw,inner sep=0.5pt] (char) {\small#1};}}

\copyrightyear{2020}
\acmYear{2020}
\setcopyright{acmlicensed}
\acmConference[LCTES '20]{21st ACM SIGPLAN/SIGBED Conference on Languages, Compilers, and Tools for Embedded Systems}{June 16, 2020}{London, United Kingdom}
\acmPrice{15.00}
\acmDOI{10.1145/3372799.3394361}
\acmISBN{978-1-4503-7094-3/20/06}

\acmArticle{1}

\title{A Collaborative Filtering Approach for the Automatic Tuning of Compiler Optimisations}

\author{Stefano Cereda}
\orcid{0000-0002-3756-3048}
\affiliation{%
  \institution{Politecnico di Milano, Milan, Italy}
}
\email{stefano.cereda@polimi.it}

\author{Gianluca Palermo}
\orcid{0000-0001-7955-8012}
\affiliation{%
  \institution{Politecnico di Milano, Milan, Italy}
}
\email{gianluca.palermo@polimi.it}

\author{Paolo Cremonesi}
\orcid{0000-0002-1253-8081}
\affiliation{%
  \institution{Politecnico di Milano, Milan, Italy}
}
\email{paolo.cremonesi@polimi.it}

\author{Stefano Doni}
\affiliation{%
  \institution{Akamas, Milan, Italy}
}
\email{stefano.doni@akamas.io}


\begin{document}

\renewcommand{\vec}[1]{\underline{#1}}

\begin{abstract}
    Selecting the right compiler optimisations has a severe impact on programs' performance.
    Still, the available optimisations keep increasing, and their effect depends on the specific program, making the
    task human intractable.
    Researchers proposed several techniques to search in the space of compiler optimisations.
    Some approaches focus on finding better search algorithms, while others try to speed up the search by leveraging
    previously collected knowledge.
    The possibility to effectively reuse previous compilation results inspired us toward the investigation
    of techniques derived from the Recommender Systems field.
    The proposed approach exploits previously collected knowledge and improves its characterisation over time.
    Differently from current state-of-the-art solutions, our approach is not based on performance
    counters but relies on Reaction Matching, an algorithm able to characterise programs looking at how they
    react to different optimisation sets.
    The proposed approach has been validated using two widely used benchmark suites, cBench and PolyBench, including 54 different programs.
    Our solution, on average, extracted 90\% of the available performance improvement 10 iterations before current state-of-the-art
    solutions, which corresponds to 40\% fewer compilations and performance tests to perform.
\end{abstract}

\maketitle

\section{Introduction}
When compiling a program from a high-level language to its executable binary, we can enable compiler optimisations
(e.g., loop unrolling, register allocation, function inlining).
These optimisations control how the code is transformed and generated, and have a severe impact on the performance of the compiled
program.
The performance can be measured as the execution time, the code size, the power consumption or the cost
efficiency, depending on the focus of interest of the user.

Compilers usually offer some predefined optimisation levels, such as
\texttt{-O1}, \texttt{-O2}, \texttt{-O3} and \texttt{-Os} for GCC\footnote{Gnu Compiler Collection
\url{https://gcc.gnu.org/}}, which are \emph{sets} containing various
optimisations empirically determined to be beneficial in most cases~\cite{hoste2008cole}.
However, the optimal set of optimisations to enable depends on the specific program to be compiled.
Such predefined sets can thus lead only to sub-optimal performance.

To find the optimal set, one could perform an exhaustive search by compiling the program with all the possible
combinations of the available optimisations and testing the program with a known and fixed workload.
However, this approach would require too much time as the search space grows exponentially with the number of available
optimisations.
GCC 9.0.1, for instance, offers 244 optimisation flags
which can be turned on or off, and 215 optimisation parameters for which we can set numerical values.
As the search space grows bigger, the need for a smart exploration strategy becomes stronger.
Researchers proposed multiple solutions to speed up this search, summarised in Section~\ref{sec:soa}.
Some focus on better exploration strategies, while others try to gain some knowledge from
previously compiled programs.

techniques derived from the field of Recommender Systems to automatically tune the compiler options.
In particular, we introduce a technique called Collaborative Filtering to suggest compiler optimisations.
This technique exploits Reaction Matching: a reaction-based approach that characterises each program not in terms
of performance counters or features, as usually done, but in terms of how different programs react to
the same optimisation set.
The resulting algorithm combines the advantages of different state-of-the-art solutions, being able to both exploit
previous knowledge and to update its belief as more compilations are performed.
We evaluate our approach, Collaborative Filtering with Reaction Matching, on the cBench and PolyBench benchmark suites.

To summarise, the proposed work introduces the following contributions:
\begin{itemize}
    \item We introduce the possibility of using techniques derived from the Recommend System field for solving the compiler autotuning problem in terms of flags selection;
    \item We combine Recommender System techniques with Reaction Matching, obtaining an approach which combines the
        advantages of different current state-of-the-art solutions and outperforms them.
\end{itemize}
The rest of the paper is organised as follows.
In Section~\ref{sec:soa} we analyse the current state of the art.
In Section~\ref{sec:methods} we give a formal description of our proposed approaches.
In Section~\ref{sec:results} we describe our experimental setup and compare different approaches.

\section{Previous work}\label{sec:soa}
Several techniques have been proposed to find good optimisation sets.
As testified by two recent and extensive surveys~\cite{ashouri2018survey, wang2018machine}, automatic tuning of compiler optimisations has recently undergone a revamp.
This is not only due to the advantages provided by a customised selection of optimisations, but also to the rise of new architectures like RISC-V (but also ARM), for which the standard compiler optimisations levels like \texttt{-O2, -O3} do not provide good-enough performance as done for Intel processors.
Instead of manually deriving novel optimisation levels, researchers focused on finding automatic ways to find good sets
of optimisations, adapting them to the particular program and target architecture.

Iterative Compilation (IC) techniques work by testing a large number of compiler optimisation sets until a sufficiently
well-performing one is found~\cite{bodin1998iterative, chen2012deconstructing}.
IC results in superior performance over predefined optimisation levels~\cite{almagor2004finding,
cooper1999optimizing}.
However, this requires long search times, which
is a significant barrier to the general adoption of these techniques.
To make IC more widespread, several search strategies have been proposed to obtain faster explorations~\cite{cooper2005acme, kulkarni2003finding, kulkarni2004fast, cooper1999optimizing, franke2005probabilistic, martins2016clustering,
Nobre:2016:GIC:2907950.2907959, fursin2010collective, Agakov:2006:UML:1121992.1122412, stephenson2003meta}.
We consider OpenTuner~\cite{ansel:pact:2014} as a representative of this family of approaches.
Instead of focusing on a specific search technique, OpenTuner contains several algorithms (such as genetic algorithms,
hill climb and multi-armed bandits) and dynamically decides which one to use.
When tuning a program, OpenTuner starts by randomly trying different techniques, and, as the tuning proceeds, it allocates
a larger proportion of tests to better performing techniques.
In this way, OpenTuner selects the best-performing search algorithm for the specific program it is tuning.
Nonetheless, many works reported that a random search performs as well as more sophisticated techniques, and is
indeed an effective tool for exploring the space of the available optimisations~\cite{Agakov:2006:UML:1121992.1122412, cavazos2007rapidly, chen2012deconstructing, kisuki2000combined}.

While search-based techniques derive their knowledge online and treat every program to be compiled as a new
search problem, other solutions try to exploit some previously-collected knowledge \cite{parello2004towards,
Agakov:2006:UML:1121992.1122412, cavazos2007rapidly, ashouri2016cobayn}.
The main idea is that the information gathered while searching the optimal set for a
program can be re-used to speed-up the search on another program.

When compiling a new program, previous knowledge is used to guide the search.
Since the knowledge base contains information about many programs, these methods need a way to understand which are the most
informative data that can be exploited for the program under compilation.
Multiple techniques can be used to identify this relevant information.

Static code features have been used in~\cite{Agakov:2006:UML:1121992.1122412}, in conjunction with machine learning
models, to guide the selection of optimisations.
The approach starts by characterising the program under compilation.
The extracted features are then transformed with Principal Component Analysis (PCA)~\cite{hotelling1933analysis}.
By using the resulting feature vector and a Euclidean distance, it is possible to find the closest program among the available ones.
This similar program is then used to focus the search process on the new program.
It is possible to use a random search (or a genetic algorithm) and speed up the search process by giving a bigger
probability of being selected to the optimisation sets which lead to performance improvement on the similar program.

This approach is refined in~\cite{cavazos2007rapidly}, where dynamically extracted features are used instead of static
ones.
Dynamic features are collected while the program runs and require costly instrumentation.
However, they can better describe the behaviour of a program.
Afterwards, logistic regression is used to directly learn a mapping from these features to the set of good
optimisations.

Cobayn~\cite{ashouri2016cobayn} also builds on this idea, combining static and dynamic features and using a Bayesian
Network (BN)~\cite{friedman1997bayesian} to focus the search process.
The proposed framework is based on a program characterisation step performed both statically on the source code using
MilePost~\cite{fursin2008milepost}
and dynamically with MICA (Microarchitecture-Independent Characterization of Applications)  metrics~\cite{hoste2007microarchitecture}.
To collect these metrics, the program is compiled with a predefined optimisation set (i.e.,
\texttt{-O3}, from now on referred to as \emph{baseline}) and run against the interesting workload.
As the program runs, the characterisation metrics are collected.
These metrics are then processed with PCA or Exploratory Factor Analysis~\cite{gorsuch1988exploratory} (EFA) to reduce their
dimensionality.
The resulting characterisation vector is fed in the BN which has been previously trained
so to output the optimal set according to the input metrics.

Most of the proposed approaches are based on the assumption that having similar characterisation metrics implies having
similar optimal sets.
However, this largely depends on the methodology used to characterise the program.
Since manually deciding which metrics may be relevant is a complex task, deep learning methodologies have been proposed
to automatically extract them from source code.
DeepTune~\cite{cummins2017end} uses a series of artificial neural networks taking as input the source code and producing
as output the expected optimal value for a configuration parameter.
Such approaches are well suited to analyse short kernel programs, but the characterization becomes harder when the code
base increases.

Feature extracted statically from the code (such as MilePost or the ones implicitly used in DeepTune) do not take into
consideration the actual workload to which the program is exposed.
Dynamic features (such as the MICA ones) avoid this problem and produce better results~\cite{ashouri2016cobayn}.
However, even dynamic features have their limitations, as they are expensive to collect and are a just a noisy proxy to
the measure we are interested in, which is how a program performs when compiled with certain optimizations.

Approaches exploting code-based features can be outperformed with a reaction-based approach:
in~\cite{cavazos2006automatic} programs are characterised in terms of the speedup they receive when compiled with
4 \emph{canonical} transformations.
By compiling and running a program with only 4 sets of optimisations they accurately
predict the speedup obtained by the program over 88000 possible sets, outperforming feature-based models.
These 4 canonical sets are decided a priori by maximising an information gain measure.

Reaction-based approaches can target both whole programs and specific code sections since they are independent
of code structure.
Moreover, by measuring performance speedups they can be tailored to a specific workload, which is not possible when
working with source code.
The approach presented in~\cite{cavazos2006automatic} is focused on the problem of predicting the speedup obtained with
a specific set, whereas we focus on a IC model, where we would like to find the optimal set as soon as possible.

FuncyTuner~\cite{wang2019funcytuner} focuses on assembling an optimised executable by separately optimising different
code regions, obtaining superior performance.
However, doing so requires to execute a greater number of compilations, which is feasible in some domains where programs
are executed repeatedly with similar inputs, but is a problem in other domains, where one can test a limited
number of binaries before the input changes.
In this work we focus on the problem of finding nearly-optimal configurations as quickly as possible, which is different
from the goal of FuncyTuner, which tries to find the best binary possible.
Nonetheless, our methodology can be combined with the per-loop optimisation strategy used in FuncyTuner to address
different domains.

MiCOMP~\cite{ashouri2017micomp} focuses on the phase-ordering problem, and uses Adjusted Cosine Similarity --- a widely used technique in the field of Recommender Systems (RS) --- as an exploration heuristic.
Considering its positive results, we are motivated to further explore the application of RS techniques to the compiler autotuning problem.

We introduce techniques derived from the RS field and combine them with a reaction-based
characterisation.
The resulting method has several interesting properties:
it can exploit past knowledge like feature-based approaches and, similarly to approaches based on dynamic features, it
is workload-dependent.
However, it is also similar to search-based techniques like OpenTuner.
Being a dynamic process, it updates its belief as more compilations are performed.
Moreover, it is extremely cheap to evaluate and it also outperforms current state-of-the-art solutions, finding
better optimisation sets in less iterations.

\section{Proposed approaches}\label{sec:methods}
In this section, we introduce the idea of using techniques inspired by the Recommender System (RS) field for the compiler
autotuning problem.

RS programs are widely used in our every-day life. They are adopted to suggest items to users, helping them to navigate
huge catalogues, like Netflix or Amazon~\cite{ricci2015recommender}.
The essential task of an RS is to predict the \emph{relevance} $r_u(i)$ of an item $i$ for a user
$u$~\cite{cremonesi2010performance}.
To achieve their goal, RS exploits similarities between items or users.

We treat the program to be tuned $p$ as our user, and a combination of the optimisation flags $x$ as the item within
the catalogue $X$, composed by all the possible sets of compiler configurations.
The relevance $r_p(x)$ reflects how much the program $p$ benefits from the optimisations defined in $x$ w.r.t. the
baseline set $x_0$ (i.e., \texttt{-O3}).
When we want to optimize a certain performance indicator (e.g., execution time), we call $f_p(x)$ the value of this
indicator obtained by the program $p$ when compiled and executed with the optimisations defined in $x$.
We consider the baseline set $x_0$ and define the relevance as:
\begin{equation}\label{eq:relevance}
    r_p(x) = \frac{f_p(x)}{f_p(x_0)} - 1
\end{equation}

To ease the search, we can exploit some previously collected information about other programs $q_0, q_1, q_2, \dots$,
where each program has been compiled and tested with a variety of optimisation sets $x_0, x_1, x_2, \dots$.
We can use RS algorithms to derive the ``preferences'' of $p$ over the available $x_s$, and then
``suggest'' to the program the optimisation sets according to the preference ordering predicted by the RS algorithm.

Here we start by presenting the simplest approach for Recommender System (RS) called Top Popular (TP)
(Section~\ref{sec:toppop}).
Then, we introduce the Content-Based Filtering (CBF) approach, which uses performance counters to characterise
programs, like most of the current state-of-the-art approaches (Section~\ref{sec:cbf}).
Finally, in Section~\ref{sec:cf}, we remove the performance counters and introduce a reaction-based characterisation
methodology called Reaction Matching (RM).
We use RM to obtain a Collaborative Filtering (CF) algorithm, which is the approach we propose as a
solution to the compiler autotuning problem.

\subsection{Top Popular --- TP}
\label{sec:toppop}
The simplest RS algorithm is the TP one, which just suggests popular
items~\cite{cremonesi2010performance}.
If most of the users have watched and liked \emph{The Lord of the Rings}, it is reasonable to suggest this item to a
new, unknown, user.
We can apply this reasoning even to programs: we should start by trying optimisation sets which are known to be
effective over most of the programs.
The underlying assumption is the same one behind the existence of standard optimisation levels: if an
optimisation set is beneficial to most programs, it is reasonable to assume that it will also be beneficial to a new
program.

The TP algorithm predicts the relevance score obtained by a program $p$ with an optimisation set $x$ as:
\begin{equation}\label{eq:rs_tp}
    \tilde{r}_p(x) = \tilde{r}(x) = \frac{\sum_{q \in Q} r_q(x)}{|Q|}
\end{equation}
that is, the relevance does not depend on the program and is predicted as the average of the relevance scores obtained by the set
$x$ on the various programs $q$ available in our knowledge base $Q$.
The knowledge base $Q=\{q_0, q_1, \dots\}$ is defined as the set of programs that we have previously explored.

\begin{figure*}
  \begin{tikzpicture}[font=\sffamily, node distance=0.5cm]
\sisetup{round-precision=2}
\node (tab1) {%
  \begin{tabular}{l|SSS}
      & \multicolumn{3}{c}{Exec time $f_q(x)$} \\
      & {$q_0$}    &   {$q_1$}   &   {$q_2$} \\
  \hline
      $x_0: \{\neg O_0, \neg O_1\}$ &   3   &   4    &   2   \\
      $x_1: \{\neg O_0, O_1\}$      &   1   &   4    &   1   \\
      $x_2: \{O_0, \neg O_1\}$      &   5   &   3    &   4   \\
      $x_3: \{O_0, O_1\}$           &   4   &   5    &   3   \\
  \end{tabular}};

  \node[right = of tab1] (tab2) {%
  \begin{tabular}{l|SSS}
      & \multicolumn{3}{c}{Relevances $r_q(x)$} \\
      & {$q_0$}    &   {$q_1$}   &   {$q_2$} \\
  \hline
      $x_0$ &   0    &   0    &   0   \\
      $x_1$ &   -0.67 &  0 &  -0.5\\
      $x_2$ &   0.67  &  -0.25 &  1\\
      $x_3$ &   0.33    &  0.25 &   0.5\\
  \end{tabular}};

  \node[right = of tab2] (tab3) {%
  \begin{tabular}{l|S}
      & {Average} \\
      & {relevances $\tilde{r}(x)$} \\
  \hline
      $x_0$ &  0     \\
      $x_1$ &  -0.39 \\
      $x_2$ &  0.47  \\
      $x_3$ &  0.36 \\
  \end{tabular}};


  \draw [line] (tab1) -- (tab2) node[midway, below, align=center] {\circled{1}};
  \draw [line] (tab2) -- (tab3) node[midway, below, align=center] {\circled{2}};

\end{tikzpicture}
  \caption{Example of TP algorithm. We have two binary optimisation flags ($O_0, O_1$), for a total of 4
    different optimisation sets ($x_0, x_1, x_2, x_3$).
    We evaluate all the sets on 3 different programs ($q_0, q_1, q_2$), compute the relevance scores and the
    average relevance across the programs.
  The TP algorithm suggests the set $x_1$ as the first one to try.
  }\label{fig:tp}
\end{figure*}

A sample execution of the TP algorithm can be found in Figure~\ref{fig:tp}.
In this example, we have 2 optimisations available $O_0$ and $O_1$, which can be combined into 4 different
optimisation sets $x_0, x_1, x_2, x_3$.
We write $x_1: \{\neg O_0, O_1\}$ to indicate that, in set $x_1$, optimisation $O_0$ is turned off and optimisation $O_1$
is turned on.
We also have 3 programs $q_0, q_1, q_2$ in our knowledge base, and we know the performance value (execution time)
obtained by every program with every optimisation set $f_q(x)$.

We start~\circled{1} by computing the relevance scores as per Equation~\ref{eq:relevance}, starting from the measured
performance values.
Then~\circled{2}, we apply Equation~\ref{eq:rs_tp}, computing the average of the relevance scores.
When tuning a fourth program $p$, the TP algorithm suggests $x_1$ as the best set, and $x_2$ as the
worst one, assuming our goal is to minimise the execution time.
Indeed, $x_1$ is the only set that reduces the execution time of all the programs in our knowledge base and thus is a
good candidate.
As we are dealing with an Iterative Compilation problem, TP proceeds by suggesting $x_0, x_2$ and $x_3$.

This algorithm is an extremely simple one and, in the RS field, is often used as a baseline algorithm.
Still, we show that it outperforms current state-of-the-art solutions, which comes at a surprise given
its simplicity.
TP can be viewed as the RS equivalent of standard optimisation levels, as it assumes that some sets are generally better
than others.
It is also similar to what one would obtain by removing the characterisation step performed in Cobayn.

\subsection{Exploiting characterisation metrics}
The TP algorithm does not require any characterisation step.
However, if we have a way to characterise programs and measure their similarity, we can apply more advanced
RS algorithms~\cite{ning2015comprehensive}.
The formula to predict relevances in this situation is:
\begin{equation}\label{eq:rs_weighted}
    \tilde{r}_p(x) = \frac
        {\sum_{q \in NN_{kp}} s_{pq}r_q(x)}
        {\sum_{q \in NN_{kp}} s_{pq}}
\end{equation}
That is, we predict the relevance of $x$ for $p$ as the weighted average of the relevance scores obtained by $x$ over the $k$
programs $q$ most similar to $p$ according to a similarity measure $s$.
We denote as $NN_{kp}$ the set of $k$ programs most similar to $p$, thus the Nearest Neighbours of $p$.

To measure the similarity between programs we take again inspiration from the RS field.
Generally speaking, an RS algorithm can belong to two broad categories: Content-Based Filtering (CBF) or
Collaborative Filtering (CF).
CBF uses items' features: if we know that a user has liked \emph{The Lord of the Rings} we can
suggest him to watch \emph{The Hobbit}, as the two items have a lot of common features (actors, genre, director,
etc) and are thus similar.
CF, conversely, does not consider features: it finds similarities between users basing on the items they liked.
If we know that two users gave similar relevance scores to many items, we can conclude that they have similar taste, and thus
consider them similar.

To obtain a CBF and a CF algorithm, we thus need to define two ways to compute similarities between programs: one based
on programs' features and one based on relevance scores.

\subsection{Content-Based Filtering --- CBF}\label{sec:cbf}
Similarly to Cobayn~\cite{ashouri2016cobayn}, we use MICA metrics~\cite{hoste2007microarchitecture} to characterise
programs and then perform a PCA~\cite{hotelling1933analysis}.
The similarity between two programs is computed using a distance metric between their feature vectors.

In Section~\ref{sec:hp_tuning} we evaluate different distance measures.
Let's suppose, for now, to use a Euclidean
distance and call $m_p^c$ the $c$-th principal component of the MICA metrics of program $p$.
Then, the distance between program $p$ and $q$ can be computed as:
\begin{equation}\label{eq:dist_metric}
    d_{pq} = \sqrt{\frac{\sum_{c=1}^C{\left(m_p^c - m_q^c \right)^2}}{C}}
\end{equation}

Since we need a similarity measure in Equation~\ref{eq:rs_weighted}, we can just define the similarity as $s_{pq} =
\frac{1}{d_{pq}}$.

This approach can be viewed as the RS equivalent of Cobayn, as it uses MICA metrics to characterise programs and then
searches for optimisation sets that are known to work well on programs with the given metrics.

\subsection{Collaborative Filtering  --- CF}\label{sec:cf}
We now introduce RM: a reaction-based characterisation methodology similar to the one proposed
in~\cite{cavazos2006automatic}.
We use RM to compute the distances in Equation~\ref{eq:rs_weighted}, obtaining a CF algorithm.

\begin{figure*}
    \input{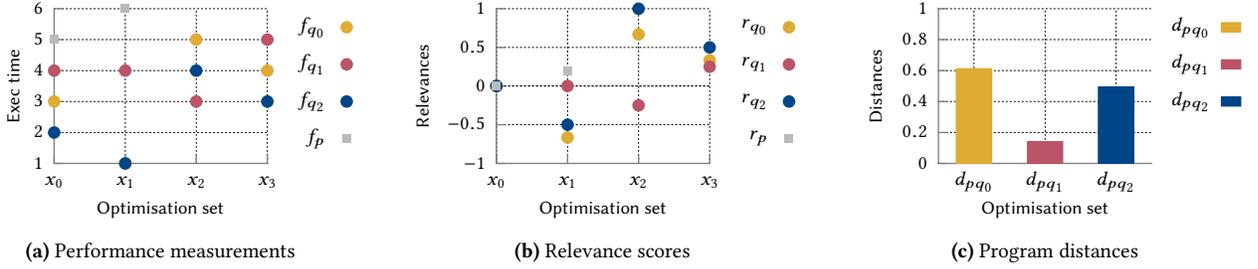}
    \caption{Example of RM algorithm used to compute program distances.
    We have three programs ($q_0, q_1, q_2$) and two optimisation flags ($O_0, O_1$) resulting in four possible
    combinations ($x_0, x_1, x_2, x_3$).
    We need to tune a fourth program $p$ for which we have only evaluated $x_0, x_1$.
    Figure~\ref{fig:norm_1} represents the value of a performance metric obtained by program $p$ when compiled with the
    flags specified in $x$, Figure~\ref{fig:norm_2} is the relevance computed with Equation~\ref{eq:relevance} and
    Figure~\ref{fig:norm_4} contains the distances between program $p$ and $q_0, q_1, q_2$ measured with
    Equation~\ref{eq:rmse} using $x_1$.
    }
    \label{fig:normalisations}
\end{figure*}

Our goal is to find a program in the knowledge base which has the same optimal set as the program we are
compiling.
If the two programs receive a performance boost from the same set of optimisations, we can hypothesise that they also
show similar reactions to other sets.
In other words, we expect the two programs to have similar code patterns so that certain optimisation sets are
beneficial to both of them, while other optimisations are detrimental, and others again have no effect at all.
To measure the similarity between two programs, we use the RM algorithm, which is graphically represented in
Figure~\ref{fig:normalisations} and described in the following.

To describe how the CF method works, we use the same initial set of data we previously used for the example in Figure~\ref{fig:tp}.
Figure~\ref{fig:norm_1} plots the measured values obtained by 3 different programs when compiled with 4 different optimisation sets.
As we did in Figure~\ref{fig:tp} for the TP algorithm, we have full information about three programs $q_0, q_1, q_2$.
Moreover, we now have a new program $p$ that needs to be tuned.
Our CF algorithm starts by evaluating the program $p$ with the baseline configuration $x_0$.
As a second configuration it evaluates the first one identified by the TP algorithm: $x_1$.
We can now start to apply RM characterisation to find the next configuration to evaluate.

In Figure~\ref{fig:norm_2} we compute the relevance scores using Equation~\ref{eq:relevance}, which brings all the baselines
to 0.
After having computed the performance scores with Equation~\ref{eq:relevance},
we define the similarity between two programs as the distance of the
relevance scores they received with the same sets.

Defining $\{x_i\}_{i=1}^n$ as the sequence of the optimisation sets explored during the $n$ iterations taken so far in the
iterative compilation of the new program $p$, and using a Euclidean distance as an example, we can compute the distance
between two programs $p, q$ as:
\begin{equation}\label{eq:rmse}
    d_{pq} = \sqrt{\frac{\sum_{i=1}^n{\left(r_p (x_i) - r_q (x_i)\right)^2}} {n}}
\end{equation}
Similarly to CBF, we obtain the similarity as $s_{pq} = \frac{1}{d_{pq}}$.

In the example of Figure~\ref{fig:norm_4}, RM uses $x_1$ to measure the distances, and identifies $q_1$ as the most
similar program to $p$.
Plugging this into Equation~\ref{eq:rs_weighted} (using $k=1$ for simplicity), RM suggests to evaluate $x_2$, which is
performing well on the similar program $q_1$.
This is in contrast with the TP algorithm, which would have suggested $x_3$ as a second candidate, and would have kept
$x_2$ as the last choice.
Once evaluated $x_2$ and having obtained $f_p(x_2)$, we need to compute its relevance score $r_p(x_2)$.
Then, we proceed by recomputing the distances and finding the next set to evaluate.

Notice that the definition  of distance depends on the optimisation sets $\{x_i\}_{i=1}^n$ which we already evaluated.
In other words, RM is an online algorithm and the computed distances vary as we proceed with IC and more sets get evaluated.
This is in contrast with TP, CBF, performance counters-based methodologies and code-based approaches which never update
their beliefs.
This puts our CF solution between solutions like Cobayn and DeepTune, which exploit previous knowledge but never
update their beliefs, and solutions like OpenTuner, which use the results of previous IC iterations to suggest the next
set to evaluate but are unable to exploit knowledge previously collected on other programs.
Moreover, by measuring the actual performance, the RM characterisation is workload-dependent and decoupled from the
source code, which can be as huge and complex as the developer likes.

\section{Experimental Evaluation}\label{sec:results}
We start by describing, in Section~\ref{sec:dataset}, the dataset on which we test our algorithms.
In Section~\ref{sec:eval_methods} we describe our evaluation methodology, which is similar to the one adopted
in~\cite{ashouri2016cobayn}.
In Section~\ref{sec:hp_tuning} we use a validation dataset to tune the algorithms.
We report our results in Section~\ref{sec:res_res} and analyze them in Section~\ref{sec:delay}.
In Section~\ref{sec:time} we compare the cost of the approaches.

\subsection{Dataset collection}\label{sec:dataset}
To validate our approach, we test it on the dataset released with Cobayn, which consists of 24 programs
taken from cBench~\cite{fursin2010collective}, each one with 5 different workloads.
We also collect another dataset using the PolyBench suite~\cite{pouchet2012polybench}.
This suite has also been used in Cobayn, but we extend the evaluation from 15 programs and 2 workloads to 30 programs
and 3 workloads.\footnote{We release the collected data at
\url{https://github.com/stefanocereda/polybench_data}}

To collect the PolyBench dataset, we use an Amazon EC2\footnote{Amazon Elastic Compute Cloud
\url{https://aws.amazon.com/ec2/}} a1.medium instance, which is equipped with a single ARMv8 gravitron processor and 2GB
of ram, Ubuntu Server 18.04 and PolyBench 4.2.1.
We collect the MICA metrics~\cite{hoste2007microarchitecture} using Intel PIN 3.10~\cite{luk2005pin}.
The cBench dataset shipped with Cobayn has instead been collected on an ARMv7 Cortex-A9 architecture as part of a
TI-OMAP 4430 processor.

In order to compare with Cobayn, we consider the 7 binary optimisations flags used
in~\cite{ashouri2016cobayn} and reported in Table~\ref{tab:opt_flags}, for a total of 128 possible different
optimisation sets, with the goal of reducing execution time.

\begin{figure}
    \begin{subfigure}{8cm}
        \input{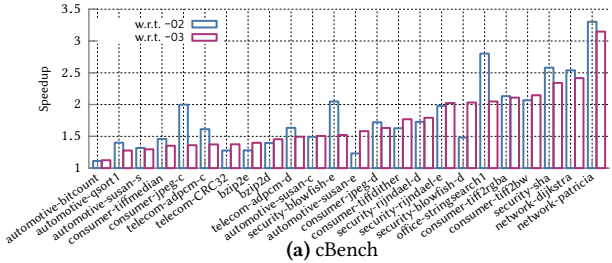}
        \subcaption{cBench}
        \label{fig:speedups_cbench}
    \end{subfigure}

    \begin{subfigure}{8cm}
        \input{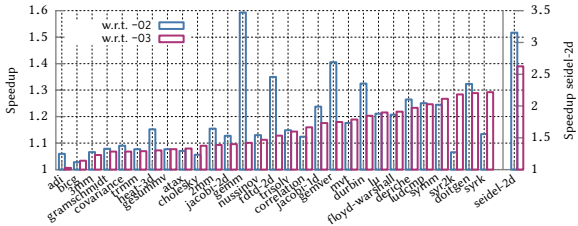}
        \subcaption{PolyBench (right axis used for seidel-2d)}
        \label{fig:speedups_polybench}
    \end{subfigure}

    \caption{Maximum performance improvement (speedup) achievable with respect to \texttt{-O2} and \texttt{-O3}. Best viewed on screen.
    }\label{fig:speedups}
\end{figure}

In Figure~\ref{fig:speedups} we report, for every program, the speedup achieved by its optimal
set w.r.t.\ \texttt{-O2} and \texttt{-O3}.

\begin{table}
    \caption{Considered optimisations.}\label{tab:opt_flags}
    \begin{tabular}{l}
        \toprule
        Optimisation\\
        \midrule
        \texttt{-funsafe-math-optimisations}\\
        \texttt{-fno-guess-branch-probability}\\
        \texttt{-fno-ivopts}\\
        \texttt{-fno-tree-loop-optimise}\\
        \texttt{-fno-inline-functions}\\
        \texttt{-funroll-all-loops}\\
        \texttt{-O2} or \texttt{-O3}\\
        \bottomrule
    \end{tabular}
\end{table}

\subsection{Evaluation Methodology}\label{sec:eval_methods}

We compare our algorithms to three state of the art approaches to iterative compilation: a Random search without
repetitions, Cobayn~\cite{ashouri2016cobayn} and OpenTuner~\cite{ansel:pact:2014}.

At the end of the experiments, we compute the \emph{optimality gap} at various iterations.
We define it as the distance from the optimal solution, measured with the Normalised Performance Improvement
(NPI) defined in~\cite{ashouri2016cobayn}:
\begin{equation}\label{eq:opt_gap}
    \text{Gap}@i = 1 - \max_{j \leq i} \left(\frac{f_p(x_0) - f_p(x_j)}{f_p(x_0) - f_p(x^{*})}\right)
\end{equation}
Where $x_0$ is the baseline set (i.e., \texttt{-O3}), $x_j$ is the optimisation set suggested by the algorithm at the
$j$th iteration, $x^*$ is the best set for program $p$ and $f_p(x_j)$ measures a certain performance indicator for program
$p$ when compiled with optimisation set $x_j$.
NPI is the argument inside the $\max$ operator and measures the ratio of the achieved performance improvement over the
available performance improvement.

Since our search algorithms start by evaluating the baseline set \texttt{-O3}, the optimality gap starts from 1, indicating that
we still have to cover all the distance to the optimum.
Conversely, as the algorithms get closer to the optimum, the gap shrinks to 0, signifying that all the potential
improvement has been achieved.
A gap of 1 thus means no performance gain over the baseline \texttt{-O3}, and a gap of 0 corresponds to the red per-column maximums in Figure~\ref{fig:speedups}.

When testing an approach, we let it work on all the available programs.
In the case of randomized approaches (i.e., Random, Cobayn and OpenTuner), we restart the algorithm from scratch and repeat the
search multiple times to reduce variability.
We repeat Cobayn and OpenTuner 10 times and Random 1000 times.
TP, CBF and CF are deterministic, so we do not repeat them.

We let every algorithm evaluate 128 optimisation sets, which comprise the entire search
space as the algorithms never repeat the same set.
When tuning a certain program, all the remaining programs (with all their workloads) are used as past knowledge, using the
same leave-one-out cross-validation approach used in~\cite{ashouri2016cobayn}.
The other workloads of the same program we are considering are not used as past knowledge.

When reporting results, we start by averaging across the repeated executions of every program, so to obtain the average
algorithm result on a specific program.
Then, we report the distributions over all the available programs and workload.

Since for some programs the baseline set is also the best one ($x_0 = x^*$), we exclude them from the evaluation
dataset.\footnote{On the cBench dataset, we removed the first two datasets of
\texttt{consumer\_jpeg\_d} program and the second one of \texttt{consumer\_tiffmedian}.
For PolyBench we removed, using a \texttt{program}-dataset notation, the following problems:
\texttt{2mm}-0,
\texttt{2mm}-2,
\texttt{3mm}-0,
\texttt{bicg}-0,
\texttt{doitgen}-1,
\texttt{durbin}-0,
\texttt{gesummv}-2,
\texttt{gramschmidt}-0,
\texttt{mvt}-0,
\texttt{symm}-0,
\texttt{syrk}-0,
\texttt{trisolv}-1,
\texttt{trmm}-0.
}
Their $NPI$ denominator would be zero, leading to an undefined optimality gap for every $i$.
This is not a problem of the method, it is only due to the NPI formulation. 

\subsection{Selection of hyperparameters}\label{sec:hp_tuning}

To implement the proposed algorithms we need to select the number of neighbours $k$, used in Equation~\ref{eq:rs_weighted},
and a distance metric, selected among Euclidean, Cosine, Chebyshev, Pearson correlation and Manhattan~\cite{cha2007comprehensive}.

To choose the hyperparameters of our algorithms we use a cross-validation approach: we randomly select two programs
(\texttt{consumer-tiff2rgba} for cBench and \texttt{mvt} for PolyBench) as our validation set, using them to select
hyperparameters and then excluding them from our test set so to avoid overfitting.
We select the hyperparameters having the lowest gap@5 on the validation set, as reported in \Cref{tab:hp_decision}.
Interestingly, we obtain similar values for the two suites.

\begin{table}
    \caption{Selected hyperparameters.}\label{tab:hp_decision}
    \footnotesize
    \begin{tabular}{lcScS}
        \toprule
        &   \multicolumn{2}{c}{CBF} & \multicolumn{2}{c}{CF} \\
        Suite   &   Metric  & $k$   &   Metric  & $k$\\
        \midrule
        cBench  &   Euclidean   &   5   &   Correlation     & 15 \\
        PolyBench  &   Euclidean   &   5   &   Correlation     & 20 \\
        \bottomrule
    \end{tabular}
\end{table}

We use the same approach for OpenTuner technique, selecting the default \emph{AUC Bandit Meta Technique A},
which is a bandit over: one differential evolution algorithm; two evolutionary techniques (uniform greedy and normal
greedy) and a simplex technique (random Nelder-Mead).
For Cobayn we use the same hyperparameters employed in~\cite{ashouri2016cobayn}.

\subsection{Experimental Results}\label{sec:res_res}
\begin{figure*}
    \begin{subfigure}{8.7cm}
        \input{./plots/cbench/self/other_apps/1.0/bsl/harmonic.tex}
        \subcaption{Harmonic average of optimality gaps on cBench.}\label{fig:plot_cbench}
    \end{subfigure}
    \addtocounter{subfigure}{6}
    \begin{subfigure}{8.7cm}
        \input{./plots/polybench/self/other_apps/1.0/bsl/harmonic.tex}
        \subcaption{Harmonic average of optimality gaps on PolyBench.}\label{fig:plot_polybench}
    \end{subfigure}

    \addtocounter{subfigure}{-7}
    \begin{subfigure}{4.34cm}
        \input{./candlesticks/cbench/self/other_apps/1.0/bsl/2.tex}
        \subcaption{Iteration \#2}\label{fig:candle_cbench_2}
    \end{subfigure}
    \begin{subfigure}{4.34cm}
        \input{./candlesticks/cbench/self/other_apps/1.0/bsl/5.tex}
        \subcaption{Iteration \#5}\label{fig:candle_cbench_5}
    \end{subfigure}
    \addtocounter{subfigure}{5}
    \begin{subfigure}{4.34cm}
        \input{./candlesticks/polybench/self/other_apps/1.0/bsl/2.tex}
        \subcaption{Iteration \#2}\label{fig:candle_polybench_2}
    \end{subfigure}
    \begin{subfigure}{4.34cm}
        \input{./candlesticks/polybench/self/other_apps/1.0/bsl/5.tex}
        \subcaption{Iteration \#5}\label{fig:candle_polybench_5}
    \end{subfigure}

    \addtocounter{subfigure}{-7}
    \begin{subfigure}{4.34cm}
        \input{./candlesticks/cbench/self/other_apps/1.0/bsl/10.tex}
        \subcaption{Iteration \#10}\label{fig:candle_cbench_10}
    \end{subfigure}
    \begin{subfigure}{4.34cm}
        \input{./candlesticks/cbench/self/other_apps/1.0/bsl/15.tex}
        \subcaption{Iteration \#15}\label{fig:candle_cbench_15}
    \end{subfigure}
    \addtocounter{subfigure}{5}
    \begin{subfigure}{4.34cm}
        \input{./candlesticks/polybench/self/other_apps/1.0/bsl/10.tex}
        \subcaption{Iteration \#10}\label{fig:candle_polybench_10}
    \end{subfigure}
    \begin{subfigure}{4.34cm}
        \input{./candlesticks/polybench/self/other_apps/1.0/bsl/15.tex}
        \subcaption{Iteration \#15}\label{fig:candle_polybench_15}
    \end{subfigure}

    \addtocounter{subfigure}{-7}
    \begin{subfigure}{4.34cm}
        \input{./candlesticks/cbench/self/other_apps/1.0/bsl/20.tex}
        \subcaption{Iteration \#20}\label{fig:candle_cbench_20}
    \end{subfigure}
    \begin{subfigure}{4.34cm}
        \input{./candlesticks/cbench/self/other_apps/1.0/bsl/25.tex}
        \subcaption{Iteration \#25}\label{fig:candle_cbench_25}
    \end{subfigure}
    \addtocounter{subfigure}{5}
    \begin{subfigure}{4.34cm}
        \input{./candlesticks/polybench/self/other_apps/1.0/bsl/20.tex}
        \subcaption{Iteration \#20}\label{fig:candle_polybench_20}
    \end{subfigure}
    \begin{subfigure}{4.34cm}
        \input{./candlesticks/polybench/self/other_apps/1.0/bsl/25.tex}
        \subcaption{Iteration \#25}\label{fig:candle_polybench_25}
    \end{subfigure}

    \caption{
        Experimental results in terms of optimality gaps on cBench
        (\cref{fig:plot_cbench,fig:candle_cbench_2,fig:candle_cbench_5,fig:candle_cbench_10,fig:candle_cbench_15,fig:candle_cbench_20,fig:candle_cbench_25})
        and PolyBench
        (\cref{fig:plot_polybench,fig:candle_polybench_2,fig:candle_polybench_5,fig:candle_polybench_10,fig:candle_polybench_15,fig:candle_polybench_20,fig:candle_polybench_25})
        reported with harmonic average (\cref{fig:plot_cbench,fig:plot_polybench}) and distributions of minimum, first
        quartile, median, third quartile and maximum over different programs
        (\cref{fig:candle_cbench_2,fig:candle_cbench_5,fig:candle_cbench_10,fig:candle_cbench_15,fig:candle_cbench_20,fig:candle_cbench_25,fig:candle_polybench_2,fig:candle_polybench_5,fig:candle_polybench_10,fig:candle_polybench_15,fig:candle_polybench_20,fig:candle_polybench_25}).
    }\label{fig:results}
\end{figure*}

In \Cref{fig:results} we report the optimality gaps obtained by the various algorithms.
Starting from cBench, in Figure~\ref{fig:plot_cbench} we aggregate over different programs of the suite using a harmonic average.
All the solutions outperform the random approach but do so in different ways. OpenTuner is very similar to random in the
first iterations, then it becomes as good as CBF and Cobayn, which, conversely, in the first iterations find better
solutions than random, suggesting that the MICA characterisation helps at finding good solutions.

However, the TP algorithm also behaves well in the first iterations, even if it does not have a characterisation step.
After 10 iterations, TP becomes significantly better than CBF and Cobayn.
CF is, most of the time, slightly below TP, suggesting that RM characterisation is effective.

To better understand the algorithms behaviour, we report in
\Cref{fig:candle_cbench_2,fig:candle_cbench_5,fig:candle_cbench_10,fig:candle_cbench_15} the quartile distributions of
the optimality gaps over different programs.
Notice that the width of the distribution does not represent the variability of algorithms, but is instead related to the
fact that some programs are harder to tune, and thus it is slower to reach small gaps.
The comparison of the variabilities of Random and TP can give us insights on the dataset nature.
A low first quartile on Random implies that there are many programs for which we can find many good optimisation sets, so it
becomes probable that Random quickly finds one of them.
A low first quartile on TP, instead, implies that there are many programs that like the \emph{same} set of flags, which
gets recommended by TP\@.
Similar reasoning can be made about the third quartile, which represents program that like few sets (Random) of very
peculiar ones (TP).
Having a low first quartile is thus easy, as it is more a property of the dataset, whereas the difficulty lies in
lowering the third quartile.
A proper characterisation should lead to a low third quartile, as it represents programs which are non-trivial to tune.

With this in mind, we can observe that the good performance of Cobayn, CBF, TP and CF in the first iterations is a more
complex story.
Looking at the second iteration¸ they all have a median gap slightly lower than the random one.
The quartiles tough are much wider, indicating that, for some programs, the characterisation is failing.
As we said earlier, the characterisation of Cobayn and CBF is fixed, so they will not be able to provide a better
characterisation for these unconventional programs.
Conversely, CF updates the RM characterisation at each iteration, and it reduces its third quartile already at
the fifth iteration.
The first quartile and the median gap of CF remain the best at all the iterations, whereas the third quartile is on par
with the one achieved by TP, but still better than the ones achieved by other approaches.

Moving to PolyBench, we report the harmonic average in \Cref{fig:plot_polybench} and the distributions in
\Cref{fig:candle_polybench_2,fig:candle_polybench_5,fig:candle_polybench_10,fig:candle_polybench_15}.

The first comment regards Random and OpenTuner.
Compared to cBench, Random is now slower while OpenTuner is faster.
This suggests that, on PolyBench, there are less good sets (which makes random slower), but they are easier to find
(which makes OpenTuner faster).
This is coherent with the behaviour of the other algorithms, which are performing much better.
Their behaviour is similar to the one we observed on cBench: Cobayn and CBF start well but, after a while, they are
reached by OpenTuner, while TP and CF performs better.
On PolyBench, Cobayn is more effective than CBF and also CF performs much better.

Looking at the quartiles we draw similar conclusions: on Polybench there are easy programs, resulting in very low first
quartiles already at the second iteration, and harder programs for which we have a high third quartile.
Even here however CF finds the best solution on all the programs already after 5 iterations, and its third quartile
remains significantly lower at all the subsequent iterations, indicating that, on PolyBench, the RM characterisation is
particularly effective.

In short, all the considered algorithms are performing better than the random one.
The second worst solution is OpenTuner, which is expected as it cannot exploit any previous knowledge.
The MICA characterisation used in CBF is not improving over a simple TP and the three best algorithms are Cobayn, TP and
CF\@.
Also notice that Cobayn substantially outperforms the random approach on both the benchmark suites, which is coherent
with the results reported in~\cite{ashouri2016cobayn} and validates the fairness of our experiments.
At the very first iterations, Cobayn and TP provide the best results on both the suites, suggesting that the good results
of Cobayn come more from implicit exploitation of the popularity bias than from an effective characterisation given by
the MICA metrics.

The main drawback of CF lies in the RM characterisation, which needs to have some points on which to base its decision
before becoming effective.
Nonetheless, in the first iterations CF is on par with TP, and, already at iteration 5, it's able to consistently find
better solutions.
Moreover, the RM characterisation becomes better and better with more iterations, letting CF keep an advantage over
other techniques even in later iterations.
RM also allows CF to have a consistently lower variability, making it a reliable algorithm even on harder-to-tune
programs.
As an example, in Figure~\ref{fig:candle_polybench_25} TP and CF look almost identical. However, the harmonic gap of CF is considerably lower, as visible in Figure~\ref{fig:plot_polybench}).

\subsection{Delay}\label{sec:delay}
To give more interpretable results, we measure the \emph{gap delay} of Cobayn, TP, CBF and OpenTuner w.r.t. CF\@.
We define the gap delay as the number of additional iterations an algorithm has to perform to reach the same
gap of CF\@.
If we want to compare an algorithm $A'$ to a reference algorithm $A$, we measure the gap delay at a certain harmonic gap
$g$ as:
\begin{equation}\label{eq:gap_delay}
        \text{gap delay}_{A', A}(g) = j' - j
\end{equation}
Where $j$ and $j'$ are the iterations needed by $A$ and $A'$ to reach an harmonic gap lower or equal than $g$.
In other words, the gap delay represents the horizontal distance between two curves in
Figures~\ref{fig:plot_cbench} and~\ref{fig:plot_polybench} measured where the reference line reaches a gap $g$.

\begin{figure*}
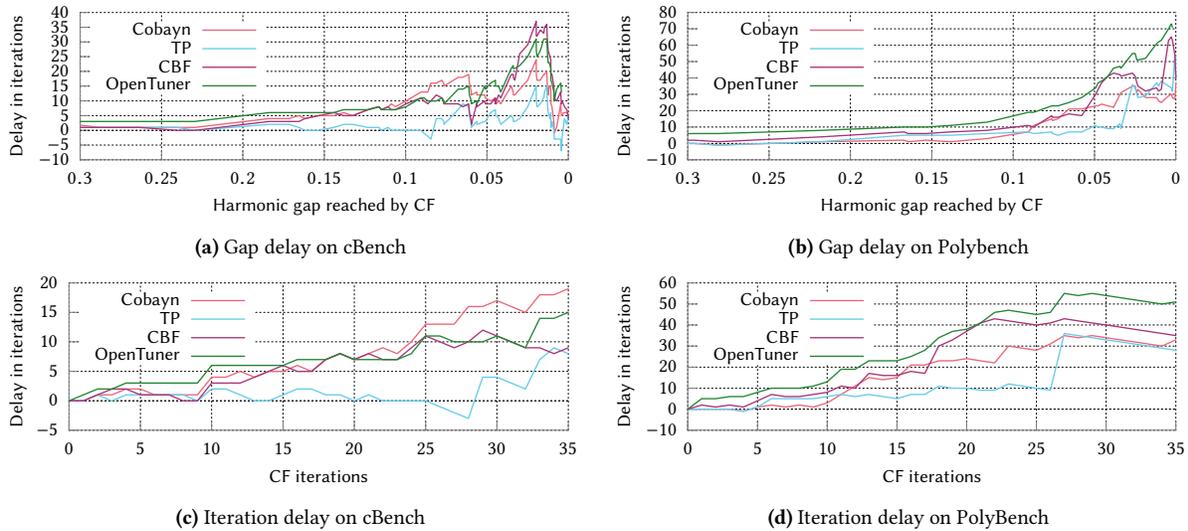

    \begin{subfigure}{8cm}
        \input{./plots_distances/plot_cbench.tex}
        \subcaption{Gap delay on cBench}\label{fig:delay_abs_cbench}
    \end{subfigure}
    \begin{subfigure}{8cm}
        \input{./plots_distances/plot_polybench.tex}
        \subcaption{Gap delay on Polybench}\label{fig:delay_abs_polybench}
    \end{subfigure}

    \begin{subfigure}{8cm}
        \input{./plots_distances_iters/plot_cbench.tex}
        \subcaption{Iteration delay on cBench}\label{fig:delay_iters_cbench}
    \end{subfigure}
    \begin{subfigure}{8cm}
        \input{./plots_distances_iters/plot_polybench.tex}
        \subcaption{Iteration delay on PolyBench}\label{fig:delay_iter_polybench}
    \end{subfigure}

        \caption{Delay of Cobayn, TP, CBF and OpenTuner wrt CF\@.
        \Cref{fig:delay_abs_cbench,fig:delay_abs_polybench} show how many additional iterations they require to obtain a
        certain gap reached by CF\@.
        \Cref{fig:delay_iters_cbench,fig:delay_iter_polybench} show how many additional iterations they need to reach
        the same gap obtained by CF after a certain number of iterations.
        }

        \label{fig:delay}
\end{figure*}

We report the gap delays in \Cref{fig:delay_abs_cbench,fig:delay_abs_polybench}.
To extract 95\% of the available performance improvement in cBench, CF has to
perform 5 compilations less than TP, 10 less than CBF and Cobayn and 15 less then OpenTuner.
On PolyBench, CF extracts 95\% performance 10 iterations before TP, 20 before Cobayn, and 30 before CBF and OpenTuner.

We also measure an \emph{iteration delay}, that is: given a fixed number of iterations and the gap reached by CF in
those iterations, how many iterations does another algorithm need to reach the same gap?
We report the results in \Cref{fig:delay_iters_cbench,fig:delay_iter_polybench}.
Even with small numbers of iterations, CF finds better solutions faster.
For instance, if we can only perform 5 iterations, CF will find solutions as good as the ones found by Cobayn in 7
iterations on cBench and 6 on PolyBench, which corresponds to 28\% and 16\% faster optimizations.
Increasing the budget to 10 iterations, the advantage of CF w.r.t. Cobayn increases to 4 and 3 iterations,
corresponding to a 28\% and 23\% faster optimizations.

In synthesis, CF is faster at finding nearly optimal solution and even in the first iterations it finds better
solutions.
With only 5 to 10 iterations, RM is already producing useful characterisations, granting an advantage to CF.

\subsection{Temporal Cost}\label{sec:time}
Here we analyse the approaches in terms of their temporal requirements.
We run all the experiments on a laptop equipped with an Intel i5-8250U CPU\@.

We start by analysing Cobayn, CBF and TP in \Cref{tab:time_cbn_cbf}.
CBF and Cobayn, in fact, require a costly MICA characterization before they can be used, which largely dominates the
cost.
Cobayn then needs to train the BN and finally use it to infer the list of suggestions, while CBF has to compute the
suggestions using \Cref{eq:dist_metric,eq:rs_weighted}.
After this initial cost, all the IC iterations are free, since COBAYN and CBF cannot update their suggestion list.
Similarly, TP uses Equation~\ref{eq:rs_tp} to compute the suggestions before starting.
\begin{table}
    \caption{Average time spent in characterization and suggestion computation for every program (measured on PolyBench).}\label{tab:time_cbn_cbf}
    \begin{tabular}{lSSS}
        \toprule
        Algorithm   &   {MICA [min]}   &     {Train [sec]}   & {Inference [sec]} \\
        \midrule
        Cobayn  &   33      &    5.65   & 1.11  \\
        CBF     &   33      &    {-}   & 0.01  \\
        TP      &   {-}     &    {-}   & 0.01  \\
        \bottomrule
    \end{tabular}
\end{table}

\begin{figure}
    \input{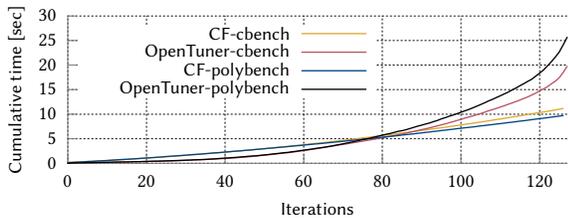}
    \caption{Time required by CF and OpenTuner.}\label{fig:time}
\end{figure}

OpenTuner and CF, instead, are inherently different, as they use the results of previous IC iterations to suggest the
next set to evaluate, and thus need to do some computations at each iteration.
We report in \Cref{fig:time} the average time required by OpenTuner and CF.
The time required by CF for a complete tuning is around 10 seconds, way below than the 33 minutes required by MICA
characterization and comparable to the time required by Cobayn if we exclude the characterization.

\section{Conclusions}\label{sec:conclusions}
We made two major contributions to the field of compiler autotuning: we introduced Recommender
Systems algorithms to select good optimisation sets and a Collaborative Filtering algorithm based on Reaction
Matching characterisation to find similar programs.

The proposed algorithm is inherently different from currently available ones, as it provides better solutions both in
the initial iterations and in later ones.
The advantage comes from the RM characterisation, which quickly finds significant similarities among programs
and becomes better and better as more sets are evaluated.

We evaluated the approaches on two benchmarking suites comprising 54 programs, considering only 7 binary
flags to compare with current state-of-the-art approaches.
However, future extensions of this work will focus on much wider sets of optimisations.
Albeit enlarging the search space will make the characterisation job harder, we are optimistic about its results,
as the approach is widely in RS, where the catalogues are much wider, and in~\cite{cavazos2006automatic} a characterisation methodology similar to our RM was able to give good characterisation by evaluating just 4 sets over
the 88000 available ones.

We have shown that the Top Popular algorithm, which has no characterisation, also outperforms current
state-of-the-art solutions, suggesting that their good result comes more from implicit exploitation of a popularity
bias than from an effective characterisation.
Conversely, the RM characterisation allows CF to outperform TP even with small numbers of evaluated flags.

CF is also an extremely cheap algorithm, as it poses no characterisation overhead over the compilation
and execution of the program and just involves a distance computation and a weighted average.
We thus think that adapting Recommender System techniques to the field of compiler autotuning is a promising path for future works.

\bibliographystyle{ACM-Reference-Format}
\bibliography{bibliography}

\end{document}